\documentclass[a4paper,11pt]{article}

\usepackage{amsthm,amsmath,amssymb,amsfonts,pstricks,pst-node,graphics}

\newcommand\A{{\mathcal A}}

\newcommand\la{{\lambda}}

\newcommand{\lieh}{{\mathfrak{h}}}

\renewcommand\d{{\rm d}}

\newcommand\cS{{\mathcal S}}

\newcommand\cH{{\mathcal H}}
\newcommand\cP{{\mathcal P}}

\def\bbbn{{\mathbb N}}

\def\bbbz{{\mathbb Z}}

\newtheorem{Def}{Definition}
\newtheorem{The}{Theorem}
\newtheorem{Pro}{Proposition}
\newtheorem{Lem}{Lemma}
\newtheorem{Ex}{Example}
\newtheorem{Cor}{Corollary}

\begin{document}
\title{Recursion operator of the Narita-Itoh-Bogoyavlensky lattice}
\author{Jing Ping Wang\\
School of Mathematics, Statistics \& Actuarial Science\\
University of Kent,
Canterbury, UK}
\date{}
\maketitle
\begin{abstract}
We construct a recursion operator for the family of Narita-Itoh-Bogoyavlensky infinite
lattice equations using its Lax presentation and present their mastersymmetries
and bi-Hamiltonian structures. We show that this highly nonlocal recursion operator 
generates infinite many local symmetries.
\end{abstract}

\section{Introduction}
Bogoyavlensky constructed in \cite{Bog88} the following two families of
integrable equations for arbitrary $p\in\mathbb{N}$ known as an integrable
discretisation for the Korteweg-de Vries equation:
\begin{eqnarray}
&&u_t=u (\sum_{k=1}^p u_{k}-\sum_{k=1}^p u_{-k}), \label{bogadd}\\
&&v_t=v (\prod _{k=1}^p v_{k}-\prod_{k=1}^p v_{-k}), \label{bogpro}
\end{eqnarray}
where both $u$ and $v$ are smooth functions of $n\in \bbbz$ and time $t$. Here
we use the notations
\begin{eqnarray*}
 u_t=\partial_t(u), \quad u_j=\cS^j u(n,t)=u(n+j,t)
\end{eqnarray*}
and $\cS$ is the shift operator. The same notation is also used for the function
$v(n,t)$ and later for the function $w(n,t)$. Equation (\ref{bogadd}) is called the
Narita-Itoh-Bogoyavlensky lattice due to the contributions of Narita and
Itoh \cite{narita, itoh}.

Taking $p=1$ in both (\ref{bogadd}) and (\ref{bogpro}), we get the well-known Volterra chain
\begin{eqnarray}\label{vol}
u_t=u (u_1-u_{-1}) .
\end{eqnarray}
Thus these two families can be regarded as the generalisation of the Volterra chain.
For fixed~$p$, equation (\ref{bogpro}) transforms into (\ref{bogadd}) under the
transformation $u= \prod_{k=0}^{p-1} v_{k}$. 

The so-called the modified Bogoyavlensky chain is given by
\begin{eqnarray}\label{bogmod}
w_t=w^2 (\prod_{k=1}^p w_{k}-\prod_{k=1}^p w_{-k}),
\end{eqnarray}
which is related to (\ref{bogadd}) by the Miura transformation $u=\prod_{k=0}^p
w_{k}$.

All systems (\ref{bogadd}), (\ref{bogpro}) and (\ref{bogmod}) can be defined
for finite lattices, that is,
the independent variable restricts to $1\leq n\leq N$ for fixed $N\in \bbbn$. 
Extensive work has been done to study
their Hamiltonian structures, associations with classical Lie algebras, the
$r$-matrix structure, etc. We refer the reader to the chapter on Bogoyavlensky
Lattices in  Suris' book \cite{suris03} including the references mentioned in
Bibliographical remarks in the end of this chapter.

In this paper, we consider these systems defined on infinite lattices, i.e.,
$n\in \bbbz$ and study their recursion operators and Hamiltonian structures
for arbitrary $p\in\bbbn$. We concentrate on system~(\ref{bogadd}). The
recursion operators and Hamiltonian operators for (\ref{bogpro}) and
(\ref{bogmod}) can be obtained via the Miura transformations mentioned before.

A recursion operator for the well-known Volterra equation (\ref{vol}) 
 is known to be 
\begin{eqnarray}\label{reV}
&&\Re=u \cS +u+u_{1} +u \cS^{-1} + u_{t} (\cS-1)^{-1} \frac{1}{u}\\
&&\quad =u(1+\cS^{-1}) (\cS u-u\cS^{-1}) (1-\cS^{-1})^{-1} u^{-1}\ .
\end{eqnarray}
It generates infinitely many local higher symmetries
\begin{eqnarray*}
&&u_{t_1}=u (u_{1}-u_{-1})\\
&&u_{t_2}=\Re (u_{t_1})=u u_{1}(u+u_{1}+u_{2})-u_{-1}u (u_{-2}+u_{-1}+u)\\
&&\cdots \cdots
\end{eqnarray*}
The Volterra equation (\ref{vol}) is a bi-Hamiltonian system
\begin{eqnarray}\label{Havol}
 u_t=\cH_1 \delta_u u=\cH_2 \delta_u \frac{\ln u}{2},
\end{eqnarray}
where $\delta_u$ is the variational derivative with respect to dependent variable
$u$,
\begin{eqnarray*}
 \cH_1=u(\cS-\cS^{-1}) u, \quad \mbox{and} \quad \cH_2=\Re \cH_1=u(1+\cS^{-1})
(\cS u-u\cS^{-1}) (1+\cS) u\ .
\end{eqnarray*}
When $p=2$, system (\ref{bogadd}) becomes
$$u_t=u(u_2+u_1-u_{-1}-u_{-2}) .
$$
Its recursion operator, mastersymmetry and bi-Hamiltonian structure were presented in \cite{mr93c:58096}, 
where the authors also gave
one Hamiltonian operator for arbitrary $p\in\bbbn$ as follows:
\begin{equation}\label{h1}
\cH=u(\sum_{i=1}^{p} \cS^i-\sum_{i=1}^{p} \cS^{-i}) u \ .
\end{equation}
Indeed, we can write system (\ref{bogadd}) as $u_t=\cH \delta_u u$. 

Recently, Svinin \cite{svin09} derived the explicit formulas of generalised symmetries for 
system (\ref{bogadd}) in terms of a family of homogeneous difference polynomials by 
considering it as a reduction of a bi-infinite sequence of the KP hierarchy. The properties of 
these homogeneous difference polynomials \cite{svin09} enable us to prove the locality of 
symmetries in Section \ref{sec4}.

In this paper, we first construct a recursion operator of (\ref{bogadd}) for
arbitrary $p$ and then show that it indeed generates local symmetries. These symmetries are the same 
as the ones presented in \cite{svin09} (modulo signs). Using the recursion operator, we further obtain the second Hamiltonian operator of (\ref{bogadd}), from which it naturally follows that the Narita-Itoh-Bogoyavlensky lattice
is a bi-Hamiltonian system.

The arrangement of the paper is as follows: In section \ref{sec2}, we give the
necessary definitions and fix the notation. The main result will appear in
section \ref{sec3}, where we give the details on how to construct the recursion
operator for all $p\in\bbbn$ via the Lax representation. In section
\ref{sec4} we prove by induction that the recursion operator 
indeed generates local symmetries . We finish the paper
with a discussion on recent results on the discrete Sawada-Kotera equation
\cite{adler1,adler2}.

\section{Basic definitions of differential-difference equations}\label{sec2}
In this section we introduce some basic concepts for differential-difference equations relevant to the contents of this paper.
More details on the variational difference complex and Lie derivatives can be found in \cite{kp85,mwx2}.

Let $u$ be a function of variable $n\in \bbbz$ and time
variable $t$. 
An evolutionary differential-difference equation of dependent variable $u$ is of the form
\begin{equation}\label{eq}
u_t=K[u], 
\end{equation} where $[u]$ means that the smooth function
$K$ depends on $u$ and its shifts. 

Assume that $K[u]$ depends on a finite set of variables $u_j$, where $q\leq j \leq p$ and $q,p \in \bbbz$
with $\partial_{u_q} K \neq 0$ and $\partial_{u_p} K \neq 0$. We say $K[u]$ is of order $(q,p)$.  
The difference between $p$ and $q$, $p-q$, is referred to as the total order of $K$. For example, 
the Volterra chain (\ref{vol}) is of order $(-1,1)$ with total order $2$.

All such smooth functions form a difference ring denoted $\A$ with the shift operator $\cS$ as its automorphism. 
For any element $f\in \A$, we define an equivalent class (or a functional) $\int\! f$
by saying that two elements $f,g\in\A$ are equivalent if \(f-g\in
\mbox{Im}(\cS-1)\).  The space of functionals is denoted by $\A'$, which
 does not inherit a ring structure of $\A$.

A vector field (derivation) ${\bf v}$ is said to be evolutionary if it commutes with
the shift operator $\cS$. Such vector filed ${\bf v}$ is completely determined by a smooth 
function $P[u]\in \A$. We call it the characteristic of vector field ${\bf v}$. We present it as
$${\bf v}=\sum_{j\in\bbbz} \cS^j(P) \frac{\partial}{\partial u_j}.
$$

For any two evolutionary vector fields with characteristics $P[u]$ and $Q[u]$, we define a Lie bracket as follows
$$[P, \ Q]=D_Q[P]-D_P[Q],
$$
where $D_{P}=\sum_j \frac{\partial P}{\partial u_j} \cS^j$ is the Fr{\'e}chet derivative of $P$. The evolutionary
vector fields form a Lie algebra denoted by $\lieh$. We simply say $P\in \lieh$.
\begin{Def}\label{def1}
An evolutionary vector field with characteristic $P[u]$ is a symmetry of system
(\ref{eq}) if and only if $[K,\ P]=0$.
\end{Def}
Equation (\ref{eq}) is said to be integrable if it possesses infinitely many higher
order symmetries. 
Often the symmetries of integrable equations can be generated by recursion operators
\cite{mr58:25341}. Roughly speaking, a recursion operator
is a linear operator $\Re:\lieh \rightarrow \lieh$ mapping a symmetry to a new symmetry.  For evolutionary equation (\ref{eq}), it satisfies
\begin{equation}\label{reopev}
D_{\Re}[K]=[ D_{K},\ \Re]\ ,
\end{equation}
where $D_{\Re}[K]$ is the Fr{\'e}chet derivative of $\Re$ along the evolutionary vector field $K$.

Symmetries of integrable systems can also be generated by mastersymmetries \cite{mr86c:58158}. An evolutionary vector
field ${\bf \tau}$ is a mastersymmetry of equation (\ref{eq}) if $[\tau, Q]$ is a new symmetry whenever $Q$ is a symmetry.
Often the vector field $\tau$ involves non-local terms.

Recursion operators for nonlinear integrable equations are often Nijenhuis
operators, that is, for any $P\in\lieh$
the operator $\Re$ satisfies
\begin{equation}\label{nijen}
D_{\Re}[\Re P]-[ D_{\Re P},\ \Re]=\Re \ (D_{\Re}[P]-[ D_{P},\ \Re])\ .
\end{equation}
Therefore, if the Nijenhuis operator $\Re$ is a recursion operator of $u_t=K$, the operator $\Re$ is also a
recursion operator for each of the evolutionary equations in the hierarchy $u_t=\Re^k
K$, where $k=0,1,2,\ldots \ .$

Nijenhuis operators are closely related to Hamiltonian and symplectic operators.
Their interrelations were discovered by Gel'fand \& Dorfman \cite{GD79,mr94j:58081} and Fuchssteiner \& Fokas
\cite{mr82g:58039,mr84j:58046}. 
The general framework 
in the context of difference variational complex and Lie derivatives can be found in \cite{kp85,mwx2}.
A generalised definition of bi-Hamiltonian systems is given in \cite{mr83d:58031,wang09}.
Here we recall the basic definitions related to Hamiltonian systems.

We denote the space of functional $q$-forms by $\Omega^q$, $q\in \bbbn$ starting with $\Omega^0=\A'$.
For any vertical 1-form $\omega=\sum_{i} h^{(i)} \d u_{i}$ on the ring $\A$, there is a
natural non-degenerate pairing with an element $P\in \lieh$:
\begin{eqnarray}\label{pairing}
<\omega, \ P>=\int \sum_{i} h^{(i)} \cS^i P = \int \left( \sum_{i} \cS^{-i} h^{(i)}
\right) \  P \ =< \sum_{i} \cS^{-i} h^{(i)},\ P>.
\end{eqnarray}
Thus the vertical 1-form $\omega$ is completely defined by $\xi \d u=\left(\sum_{i} \cS^{-i} h^{(i)}\right) \d u$. 
We simply say $\xi\in \Omega^1$.

The pairing (\ref{pairing}) allows us to give the definition of (formal) adjoint
operators to linear (pseudo-) difference operators \cite{kp85,mwx2}.
\begin{Def} Given a linear operator ${\cal A}: \lieh \rightarrow  \Omega^1$,
we call the operator ${\cal A}^{\dagger}: \lieh \rightarrow \Omega^1$ the adjoint
operator of ${\cal A}$ if $<{\cal A} P, \ Q>=<{\cal A}^{\dagger} Q, \ P>$,
where $P, Q\in \lieh$.
\end{Def}
Similarly, we can define the adjoint operator for an operator mapping from $\Omega^1$
to $\lieh$, from $\lieh$ to $\lieh$ or from $\Omega^1$ to $\Omega^1$. We say an operator $A$ is anti-symmetric if 
$A^{\dagger}=-A$.

For any functional $\int\!\!g\in \A'$, we define its difference variational derivative (Euler operator) denoted by
$\delta_{u} (\int\!\! g) \in \Omega^1$ as
$$\delta_{u} (\int\! g)=\sum_{i\in\bbbz} \cS^{-i}  \frac{\partial g}{\partial u_{i}}=
\frac{\partial }{\partial u}\left(\sum_{i\in\bbbz} \cS^{-i}  g \right)\in \Omega^1.$$
\begin{Def}
An anti-symmetric operator $\cH: \Omega^1\rightarrow \lieh$ is Hamiltonian if and only if the bracket defined on $\A'$
as follows:
$$
\left\{ \int\!\! f, \ \int\!\! g\right\}=\int\!\! \left(\delta_u f\ \cH \delta_u g\right)
$$
satisfies the Jacobi identity.
\end{Def}

\begin{Def}
We say a Hamiltonian operator $\cH: \Omega^1\rightarrow \lieh$ is a Hamiltonian operator for equation (\ref{eq})
if its Lie derivative along evolutionary vector field $K\in \lieh$ vanishes, that is
\begin{eqnarray}\label{lieH}
L_K\! \cH:=D_{\cH}[K] -D_K \cH-\cH D_K^{\dagger}=0. 
\end{eqnarray}
\end{Def}
Alternatively, we can define Hamiltonian operators in term of the Schouten bracket.
\begin{Def}
Let $\cH,\cP:\Omega^1\rightarrow \lieh$ be two anti-symmetric operators. The Schouten bracket of
$\cH$ and $\cP$ is the trilinear mapping 
$[[\cH,\ \cP]]:\Omega^1\times\Omega^1\times\Omega^1\rightarrow \Omega^0$ defined by
\begin{eqnarray}\label{schouten}
\begin{array}{l}
[[\cH,\ \cP]](\xi_1,\xi_2,\xi_3):=<\xi_1, D_{\cH}[\cP \xi_3](\xi_2)>+<\xi_1, D_{\cP}[\cH \xi_3](\xi_2)>
+<\xi_2, D_{\cH}[\cP \xi_1](\xi_3)>\\
\quad+<\xi_2, D_{\cP}[\cH \xi_1] (\xi_3)>+<\xi_3, D_{\cH}[\cP \xi_2](\xi_1)>
+<\xi_3, D_{\cP}[\cH \xi_2](\xi_1)>.
\end{array}
\end{eqnarray}
\end{Def}
By direct calculation, it can be shown that an antisymmetric operator $\cH$ is Hamiltonian if and only if
$[[\cH,\ \cH]]=0$, \cite{mr94j:58081,mr94g:58260}. Moreover, the relation between Nijenhuis operators 
and Schouten brackets can be formulated as follows:
\begin{The} Let $\cH_1, \cH_2$ be two anti-symmetric operators and $\Re=\cH_1 \cH_2^{-1}$. Then
 \begin{eqnarray}\label{NHR}
 \begin{array}{l}
< \xi, D_{\Re}[\Re P](Q)-\Re D_{\Re}[P] (Q)-D_{\Re}[\Re Q](P)+\Re D_{\Re}[Q](P)>\\
=\frac{1}{2} [[\cH_1,\ \cH_1]](\xi,\xi_1,\xi_2)+\frac{1}{2} [[\cH_2,\ \cH_2]](\Re^{\dagger 2}\xi,\xi_1,\xi_2)
-[[\cH_1,\ \cH_2]](\Re^{\dagger}\xi,\xi_1,\xi_2),
 \end{array}
\end{eqnarray}
where $P=\cH_2 \xi_1$ and  $Q=\cH_2 \xi_2$.
\end{The}
It follows that operator $\Re=\cH_1 \cH_2^{-1}$ is Nijenhuis if $\cH_1$ and $\cH_2$ form a Hamiltonian pair.
\begin{Def}
An evolutionary equation (\ref{eq}) is said to be bi-Hamiltonian if there exist two Hamiltonian operators $\cH_0$
and $\cH_1$, and two Hamiltonians $f_0$ and $f_1$ such that
\begin{equation*}
 u_t=\cH_0 \delta_u f_0=\cH_1 \delta_u f_1 ,
\end{equation*}
where $\delta_u$  is the difference variational derivative with respect to $u$.
\end{Def}
\section{Construction of recursion operators from a Lax representation}\label{sec3}
In general, it is not easy to construct a recursion operator for a
given integrable equation although we have the explicit formula (\ref{reopev}). The difficulty lies in how 
to determine the starting
terms of $\Re$, i.e., the order of the operator, and how to construct its
nonlocal terms. Many papers are devoted to this subject, see
\cite{mr88g:58080,mr1974732,gw11}. 

If the Lax representation of an evolutionary equation is
known, an amazingly simple approach to
construct a recursion operator was proposed in \cite{mr2001h:37146} and later applied for lattice equations
 \cite{Maciej200127}. 
This idea can be used for Lax pairs that are
invariant under the reduction groups \cite{wang09}. 
In this section, we use the same idea to construct a recursion operator of system (\ref{bogadd}) from
its Lax representation. The recursion operators of (\ref{bogpro}) and
(\ref{bogmod}) can then be derived via the Miura transformations.

The Lax operator of (\ref{bogadd}) for any $p\in\bbbn$ is given in \cite{Bog88} as follows:
\begin{equation}\label{lax0}
L=\cS+u\cS^{-p}\ . 
\end{equation}
We can compute its hierarchy of symmetry flows using the formula
\begin{eqnarray*}
&&L_{t_n}=[B^{(n)},\ L], \qquad B^{(n)}=(L^{(p+1)n})_{\geq 0},
\end{eqnarray*}
where $\geq 0$ means taking the terms with non-negative power of $\cS$ in $L^{(p+1)n}$.

The idea to construct a recursion operator directly from a Lax representation is
to relate two difference operators such as $B^{(n_1)}$ and $B^{(n_2)}$
and to find the relation between the corresponding flows
\cite{mr2001h:37146}. 

We rewrite (\ref{lax0}) into the following matrix form
\begin{equation}\label{lax}
 L(\la)=\cS-\la U^{(0)} - U^{(1)}:=\cS-C(\la) ,
\end{equation}
where $U^{(0)}=(u^{(0)}_{i,j})$ and $U^{(1)}=(u^{(1)}_{i,j})$ are $(p+1)\times
(p+1)$ matrices.  The matrices $U^{(0)}$ and $U^{(1)}$ are given by
\begin{eqnarray*}
U^{(0)}  =\left(\begin{array}{lcrl} 1 &0 & \cdots&  0 \\ 0 & 0 & \cdots
& 0\\
\vdots& \vdots & \ddots &\vdots \\0&0&\cdots&0 \\0&0&\cdots&  0
\end{array} \right) 
\quad \mbox{and}\quad
U^{(1)}  =\left(\begin{array}{lcrll} 0 &0 & \cdots& 0 & -u \\ 1 & 0 & \cdots & 0
& 0\\
\vdots & \ddots& &\vdots&\vdots \\0&\cdots&1&0&0 \\0&\cdots& 0& 1&0
\end{array} \right) 
\end{eqnarray*}
respectively, that is, their non-zero entries are
$$ u^{(0)}_{11}=1, \quad u^{(1)}_{1,p+1}=-u \quad \mbox{and}\quad u^{(1)}_{i+1,i}=1, \ \ i=1,2,\cdots,
p.$$
Further we rewrite the operator $B$ in a matrix form denoted by $B(\la)$. So the symmetry flows can be obtained by the zero curvature equation
\begin{equation}\label{zero}
 C(\la)_t=\cS(B(\la))\ C(\la)-C(\la) B(\la) .
\end{equation}
Make the Ansatz
\begin{eqnarray}\label{mw}
\bar B(\la)=\la^{p+1} B(\la)+ W(\la),\quad \mbox{where} \quad W(\la)=\sum_{i=0}^{p+1}\la^{p+1-i}A^{(i)},
\end{eqnarray}
where $A^{(i)}=(a^{(i)}_{kl})$ are $(p+1)\times (p+1)$ matrices 
with the only non-zero entries being $a^{(i)}_{j+i,j}$ for $1\leq j\leq p+1$.
Here we read $i+j$ as $(i+j) \mod (p+1)$. For simplicity, we shall continue to
denote the index $l$ when $l>p+1$ instead of $l\mod (p+1)$. So both $A^{(0)}$ and $A^{(p+1)}$ are diagonal 
matrices.

The Ansatz $W$ is invariant under the following transformation
$$
r: W(\la)\mapsto P W(\sigma \la) P^{-1} ,
$$
where $P$ is a diagonal $(p+1)\times (p+1)$ matrix given by $P_{ii}=\sigma^{i}$ and
$\sigma=e^{2\pi i/(p+1)}$ since we have $P^{-1} A^{(i)} P=\sigma^i A^{(i)}$. The transformation satisfies
$r^{p+1} = id $
and therefore generates the group $\bbbz_{p+1}$. The reduction groups of Lax
pairs have been studied in \cite{mik80,mik81}.

The zero curvature condition (\ref{zero}) for $\bar B(\la)$ leads to
a formula for computing a recursion operator as follows:
\begin{eqnarray}\label{rel}
C_{t_{n+1}}=\cS(\bar B(\la))\ C(\la)-C(\la) \bar B(\la) =\la^{p+1} C_{t_n} + \cS(W) C-C W.
\end{eqnarray}
Substituting (\ref{lax}) and (\ref{mw}) into (\ref{rel}) and collecting the
coefficients of this $\la$-polynomial, we obtain
\begin{eqnarray}
&&\la^{p+2}:\quad     \cS(A^{(0)}) U^{(0)}-U^{(0)} A^{(0)} =0; \label{a0}\\
&&\la^{p+1}: \quad   U^{(1)} _{t_n}+\cS(A^{(1)}) U^{(0)}-U^{(0)}
A^{(1)}+\cS(A^{(0)})
U^{(1)}-U^{(1)} A^{(0)}=0;\label{a1}\\
&&\la^{p+1-i}: \   \cS(A^{(i+1)}) U^{(0)}-U^{(0)} A^{(i+1)}+\cS(A^{(i)})
U^{(1)}-U^{(1)} A^{(i)}=0, \quad 1\leq i\leq p;\label{ai}\\
&&\la^{0}: \qquad    U^{(1)}_{t_{n+1}}=\cS(A^{(p+1)}) U^{(1)}-U^{(1)}
A^{(p+1)}.\label{ap}
\end{eqnarray}
\begin{Lem}\label{lem1}
Assume that $p\geq 2$ and $ 1\leq i\leq p-1$. The matrix equations 
\begin{equation}\label{matrixeq}
\cS(A^{(i+1)}) U^{(0)}-U^{(0)} A^{(i+1)}+\cS(A^{(i)}) U^{(1)}-U^{(1)} A^{(i)}=0 
\end{equation}
are equivalent to 
\begin{eqnarray}\label{reai}
&&\cS(a^{(i+1)}_{i+2,1})+\cS(a^{(i)}_{i+2,2})-a_{1+i,1}^{(i)}=0
;\label{reai1}\\
&&-a^{(i+1)}_{1,p+1-i}+\cS(a^{(i)}_{1,p+2-i})+u
a_{p+1,p+1-i}^{(i)}=0;\label{reai2}\\
&&-u \cS(a^{(i)}_{i+1,1})-a_{i,p+1}^{(i)}=0;\label{reai3}\\
&&\cS(a^{(i)}_{l+i+1,l+1})-a_{l+i,l}^{(i)}=0,\quad 2\leq l\leq
p,\quad l\neq p+1-i \ .\label{reai4}
\end{eqnarray}
\end{Lem}
\noindent
{\bf Proof}. We directly compute the multiplications of matrices and write
out their non-zero entries, respectively:
\begin{eqnarray*}
&&( \cS(A^{(i+1)}) U^{(0)})_{i+2,1}=\cS(a^{(i+1)}_{i+2,1})
u^{(0)}_{11}=\cS(a^{(i+1)}_{i+2,1});\\
&&(U^{(0)} A^{(i+1)})_{1,p+1-i}=u^{(0)}_{11}
a^{(i+1)}_{1,p+1-i}=a^{(i+1)}_{1,p+1-i};\\
&&(\cS(A^{(i)}) U^{(1)})_{i+1,p+1}=\cS(a^{(i)}_{i+1,1}) u_{1,p+1}^{(1)}=-u
\cS(a^{(i)}_{i+1,1});\\
&&(\cS(A^{(i)}) U^{(1)})_{l+i+1,l}=\cS(a^{(i)}_{l+i+1,l+1})
u_{l+1,l}^{(1)}=\cS(a^{(i)}_{l+i+1,l+1}),\quad 1\leq l\leq p;\\
&&(U^{(1)} A^{(i)})_{1,p+1-i}=u_{1,p+1}^{(1)} a_{p+1,p+1-i}^{(i)}=-u
a_{p+1,p+1-i}^{(i)} ;\\
&&(U^{(1)} A^{(i)})_{l+i+1,l}=u_{l+i+1,l+i}^{(1)}
a_{l+i,l}^{(i)}=a_{l+i,l}^{(i)},\quad 1\leq l\leq p+1, \quad l\neq p+1-i;
\end{eqnarray*}
We are now ready to write out the entries for the matrix equations, which leads to the formulas
stated in the lemma.
\hfill $\blacksquare$

Specifically, when $i=0$, using the proof of Lemma \ref{lem1} we obtain the equivalent conditions for 
the matrix equation (\ref{matrixeq}) as follows:
\begin{eqnarray}
&&\cS(a^{(1)}_{2,1})+\cS(a^{(0)}_{2,2})-a_{1,1}^{(0)}=0;
\label{rea00}\\
&&-a^{(1)}_{1,p+1}+\cS(a^{(0)}_{1,1})+u
a_{p+1,p+1}^{(0)}=0;\label{rea01}\\
&&\cS(a^{(0)}_{l+1,l+1})-a_{l,l}^{(0)}=0,\quad 2\leq l\leq
p \ .\label{rea02}
\end{eqnarray}
In the similar way, the matrix equation (\ref{matrixeq}) for $i=p$ is equivalent to
\begin{eqnarray}
&&\cS(a^{(p+1)}_{1,1})-a^{(p+1)}_{1,1}+\cS(a^{(p)}_{1,2})+u
a_{p+1,1}^{(p)}=0;\label{reap0}\\
&&-u \cS(a^{(p)}_{p+1,1})-a_{p,p+1}^{(p)}=0;\label{reap1}\\
&&\cS(a^{(p)}_{l,l+1})-a_{l-1,l}^{(p)}=0,\quad 2\leq l\leq p \ .\label{reap2}
\end{eqnarray}
Notice that formula (\ref{rea00}), (\ref{rea01}), (\ref{reap0}) and (\ref{reap1}) are valid for $p=1$.

Using formula (\ref{reai})--(\ref{reai4}) in Lemma \ref{lem1}, we can now find the relation between $a^{(i+1)}_{i+2,1}$ 
and $a^{(i)}_{i+1,1}$.
\begin{Lem}\label{lem2}
Assume that $p\geq 2$ and $1\leq i \leq p-1$. We have
\begin{eqnarray}\label{keyre}
a^{(i+1)}_{i+2,1}=\cS^{-1} (\cS^{i} u-u \cS^{i-p})^{-1} (\cS^{i} u  -u
\cS^{i-p-1}) \cS  (a^{(i)}_{i+1,1}) .
\end{eqnarray}
\end{Lem}
\noindent
{\bf Proof}. First using formula (\ref{reai4}), we can show that 
\begin{eqnarray}\label{a1pi}
a_{1,p+2-i}^{(i)}=\cS^{i-1} (a_{i,p+1}^{(i)}) .
\end{eqnarray}
Indeed, if we take $l=p+2-i+r$ in (\ref{reai4}), it follows that
$$a_{1+r,p+2-i+r}^{(i)}=\cS(a^{(i)}_{2+r,p+3-i+r}), \quad 0\leq r \leq i-2 .
$$
Thus we recursively obtain (\ref{a1pi}), that is,
$$a_{1,p+2-i}^{(i)}=\cS(a^{(i)}_{2,p+3-i})=\cdots =\cS^{i-1} (a_{i,p+1}^{(i)}) .
$$ 
From (\ref{reai3}), it leads to $a_{i,p+1}^{(i)}=-u \cS(a^{(i)}_{i+1,1})$. Substituting it into to (\ref{a1pi}), we get 
\begin{eqnarray}\label{sai}
a_{1,p+2-i}^{(i)} =-\cS^{i-1} u \cS(a^{(i)}_{i+1,1}). 
\end{eqnarray}
Similar as the proof of (\ref{a1pi}), by taking $ l=2+r$, where $0\leq r\leq p-2-i$ in (\ref{reai4}) when $p\geq 2$, we recursively show that
\begin{eqnarray}\label{saip}
a_{2+i,2}^{(i)} =\cS(a^{(i)}_{3+i,3})=\cdots =\cS^{p-1-i}
(a_{p+1,p+1-i}^{(i)}).
\end{eqnarray}
Now we substitute (\ref{saip}) into (\ref{reai1}) and it leads to
\begin{eqnarray}\label{pi}
\cS(a^{(i+1)}_{i+2,1})+\cS^{p-i} (a_{p+1,p+1-i}^{(i)})-a_{1+i,1}^{(i)}=0.
\end{eqnarray}
Formula (\ref{sai}) is valid for different value of $1\leq i\leq p-1$. Thus we have
$a_{1,p+1-i}^{(i+1)} =-\cS^{i} u \cS(a^{(i+1)}_{i+2,1})$.  We substitute it and (\ref{sai}) into (\ref{reai2}) and it becomes 
\begin{eqnarray}\label{pii}
 \cS^{i} u \cS(a^{(i+1)}_{i+2,1})-\cS^{i} u
\cS(a^{(i)}_{i+1,1})+u a_{p+1,p+1-i}^{(i)}=0 .
\end{eqnarray}
From (\ref{pi}) and (\ref{pii}), we eliminate $a_{p+1,p+1-i}^{(i)}$ and obtain
the relation between $a^{(i+1)}_{i+2,1}$ and $a^{(i)}_{i+1,1}$, that is,
\begin{eqnarray*}
a^{(i+1)}_{i+2,1}=(\cS^{i} u \cS-u \cS^{i+1-p})^{-1} (\cS^{i} u \cS -u
\cS^{i-p})   (a^{(i)}_{i+1,1}),
\end{eqnarray*}
which is equivalent to (\ref{keyre}) as written in this lemma.
\hfill $\blacksquare$

From (\ref{reap1}) and (\ref{reap2}), we can see that (\ref{sai}) is also valid  for $i=p$. Substituting it into
(\ref{reap0}), we obtain 
$$(\cS-1)a^{(p+1)}_{1,1}=-\cS(a^{(p)}_{1,2})-u a_{p+1,1}^{(p)} =(\cS^{p} u
\cS -u) (a^{(p)}_{p+1,1}).
$$
Therefore,
\begin{eqnarray}\label{keyp}
a^{(p+1)}_{1,1}=(\cS-1)^{-1} (\cS^{p} u -u\cS^{-1})\cS (a^{(p)}_{p+1,1}).
\end{eqnarray}

We are now ready to compute the recursion operators for any $p\in\bbbn$.
\begin{The}\label{th1}
A recursion operator of Narita-Itoh-Bogoyavlensky lattice 
(\ref{bogadd}) is
 \begin{eqnarray}\label{Re}
\Re=u (\cS -\cS^{-p}) (\cS -1)^{-1} \prod_{i=1}^{\rightarrow{p}} (\cS^{p+1-i} u-u\cS^{-i} )
(\cS^{p-i} u-u \cS^{-i})^{-1} \ .
\end{eqnarray}
\end{The}
Since the difference operators are not commuting, here we use the notation 
$\prod_{i=1}^{\rightarrow{p}}$ to denote the order of the value $i$, from $1$ to $p$, that is,
$\prod_{i=1}^{\rightarrow{p}}a_i=a_1 a_2 \cdots a_p$.

\noindent
{\bf Proof}. First from (\ref{rea02}), we can show that
\begin{eqnarray}\label{A0}
a^{(0)}_{ll}=\cS^{p+1-l}(a^{(0)}_{p+1,p+1}),\quad 2\leq l \leq p+1 .
\end{eqnarray}
The next identity (\ref{a0}) leads to $(\cS-1)a^{(0)}_{11}=0$. Here we choose the solution
$a^{(0)}_{11}=0$, which makes it possible to find the relation between $u_{t_{n+1}}$ and $u_{t_n}$ .
We now substitute this into (\ref{rea00}) and using (\ref{A0}) we obtain
\begin{eqnarray}\label{A1p1}
a^{(1)}_{2,1}=-a^{(0)}_{2,2}=-\cS^{p-1}(a^{(0)}_{p+1,p+1}) .
\end{eqnarray}
It follows from (\ref{a1}) and (\ref{rea01}) that
\begin{eqnarray*}
-u_{t_n}-a^{(1)}_{1,p+1}+u a_{p+1,p+1}^{(0)}=0 .
\end{eqnarray*}
We now show that $a^{(1)}_{1,p+1}=-u \cS (a^{(1)}_{2,1})$ for all $p\in\bbbn$. When $p\geq 2$, 
we have $a^{(1)}_{1,p+1}=-u \cS (a^{(1)}_{2,1})$ by taking $i=1$ in (\ref{reai3}). With $p=1$, this is a result from  (\ref{reap1}). Therefore for all $p\in \bbbn$, using (\ref{A1p1}) we have
 $a^{(1)}_{1,p+1}=u \cS^p (a^{(0)}_{p+1,p+1})$ . This leads to
\begin{eqnarray*}
u_{t_n}+u(\cS^{p}-1) a_{p+1,p+1}^{(0)}=0.
\end{eqnarray*}
Therefore, we have
\begin{eqnarray}\label{ret}
a_{p+1,p+1}^{(0)}=(1-\cS^{p})^{-1} \frac{u_{t_n}}{u} \ .
\end{eqnarray}
Form (\ref{ap}), we find that
\begin{eqnarray*}
\left\{\begin{array}{l}
-u_{t_{n+1}}=-u\cS(a^{(p+1)}_{1,1})+u a_{p+1,p+1}^{(p+1)};\\
\cS(a^{(p+1)}_{l+1,l+1})=a_{l,l}^{(p+1)},\quad 1\leq l\leq p \ .\end{array}\right.
\end{eqnarray*}
This implies that
\begin{eqnarray*}
 u_{t_{n+1}}=u (\cS -\cS^{-p}) (a^{(p+1)}_{1,1}).
\end{eqnarray*}
Using (\ref{keyp}) we rewrite it as
\begin{eqnarray*}
 u_{t_{n+1}}=u (\cS -\cS^{-p}) (\cS-1)^{-1} (\cS^{p} u -u\cS^{-1} )\cS
(a^{(p)}_{p+1,1}). 
\end{eqnarray*}
Using Lemma \ref{lem2} when $p\geq 2$, we obtain
\begin{eqnarray*}
 u_{t_{n+1}}=u (\cS\! -\!\cS^{-p}) (\cS\!-\!1)^{-1} (\cS^{p} u\! -\!u\cS^{-1})
\prod_{i=1}^{\rightarrow{(p-1)}} (\cS^{p-i} u\!-\!u \cS^{-i})^{-1} (\cS^{p-i} u \! -\!u
\cS^{-i-1}) \cS (a^{(1)}_{2,1})\ .
\end{eqnarray*}
This is also valid for $p=1$ under the convention that the empty product is equal to $1$.
Finally, using (\ref{A1p1}) and (\ref{ret}), we get the relation between
$u_{t_{n+1}}$ and $u_{t_n}$, which gives rise to the recursion operator as
stated in this theorem.
\hfill $\blacksquare$

By directly checking the identity (\ref{nijen}), we can prove that the operator $\Re$ 
in Theorem \ref{th1} is Nijenhuis. Here we omit the proof since we have not found a neat way 
to present it in reasonable length even for a given $p$,

Notice that equation (\ref{bogadd}) admits a scaling symmetry $u$, that is, $[u,\ u_t]=u_t$.
Knowing the recursion operator, one can obtain its master symmetry \cite{mr86c:58158}.
\begin{Cor}\label{cor1} A master symmetry for equation (\ref{bogadd}) is $\tau=\Re u$, where $\Re$ is the recursion operator given by
(\ref{Re}) in Theorem \ref{th1}.
\end{Cor}
\begin{Ex}
For $p=1$ in (\ref{bogadd}, it becomes the Volterra chain (\ref{vol}). From Theorem \ref{th1}, we get the same recursion operator 
as given by (\ref{reV}). It follows from Corollary \ref{cor1}, its mastersymmetry is
$$\tau=\Re_{p=1} u=n u (u_1-u_{-1}) +u (u_1 +u+2 u_{-1}) .
$$
The Lie derivative of its Hamiltonian operator $\cH_1$ defined by (\ref{Havol}) along vector field $\tau$ leads to 
\begin{eqnarray*}
L_{\tau} \cH_1=D_{\cH_1}[\tau]-D_{\tau} \cH_1-\cH_1 D_{\tau}^{\dagger}=-\cH_2,
\end{eqnarray*}
which is defined by (\ref{Havol}). It is also a Hamiltonian operator of the Volterra chain.
\end{Ex}


For equation (\ref{bogadd}), one Hamiltonian operator $\cH$ given by (\ref{h1})
is known \cite{mr93c:58096}. We now compute $\hat \cH=\Re \cH$, that is,
\begin{eqnarray*}
\hat \cH=u (\sum_{i=0}^p \cS^{-i}) \left(
\prod_{i=1}^{\rightarrow{(p-1)}}(\cS^{p+1-i} u-u\cS^{-i} ) (\cS^{p-i}
u-u \cS^{-i})^{-1} \right) (\cS u-u\cS^{-p} )
\left(\sum_{i=0}^p \cS^{i}\right) u\ ,
\end{eqnarray*}
This operator is anti-symmetric since it satisfies ${\hat \cH}^\dagger=-\hat \cH$.
\begin{Pro}
The above anti-symmetric operator $\hat \cH$ is Hamiltonian.
\end{Pro}
\noindent
{\bf Proof}. Here we sketch the proof. First by direct calculation we have $L_{\tau} \cH=-\hat \cH$.
Since $\cH$ is Hamiltonian, its Schouten bracket vanishes, that is, $[[\cH, \ \cH]]=0$.
It implies that $[[L_{\tau} \cH,\ \cH]]=0$, \cite{mr94j:58081}, implying $[[\hat \cH,\ \cH]]=0$.
Due to $\Re$ being Nijenhuis, it follows from (\ref{NHR}) the Schouten bracket of $\hat \cH$ vanishes. Thus it is a Hamiltonian
operator.
\hfill $\blacksquare$

Hence the Narita-Itoh-Bogoyavlensky lattice is a bi-Hamiltonian system. Indeed,
we can write equation (\ref{bogadd}) as
\begin{eqnarray}\label{bih}
u_t=\cH \delta_u  u=\Re \cH \delta_u \frac{\ln u}{p+1} \ . 
\end{eqnarray}
The recursion operators and Hamiltonian operators for equations (\ref{bogpro})
and (\ref{bogmod}) can be obtained by the Miura transformations. Notice that
\begin{eqnarray*}
&&D_u= u \left(\sum_{k=0}^{p-1} \frac{1}{v_k} \cS^k\right)=u
\left(\sum_{k=0}^{p-1}  \cS^k\right) \frac{1}{v}
=u \left(\sum_{k=0}^{p} \frac{1}{w_k} \cS^k\right)=u
\left(\sum_{k=0}^{p}  \cS^k\right) \frac{1}{w} .
\end{eqnarray*}
Using Theorem \ref{th1}, we obtain the following results:
\begin{Cor}
For any $p\in\bbbn$, a recursion operator of equation (\ref{bogpro}) is
 \begin{eqnarray}
&&\Re_v=v (\cS^p -1)^{-1}(\cS -\cS^{-p}) \cdot  \prod_{i=1}^{\rightarrow{p-1}}
(\cS^{p+1-i} u-u\cS^{-i} )(\cS^{p-i} u-u \cS^{-i})^{-1} \nonumber\\
&&\qquad \cdot (\cS u-u\cS^{-p} ) \cS^p ( v \cS -v)^{-1} , \label{Rev}
\end{eqnarray} 
where $u=\prod_{k=0}^{p-1} v_{k}$.
A recursion operator of equation (\ref{bogmod}) is
 \begin{eqnarray}
&&\Re_w=w \cS^{-p} \cdot  \prod_{i=1}^{\rightarrow{p-1}}
(\cS^{p+1-i} u-u\cS^{-i} )(\cS^{p-i} u-u \cS^{-i})^{-1} \nonumber\\
&&\qquad \cdot (\cS u-u\cS^{-p} ) (1-\cS^{-p})^{-1} (\cS^{p+1}-1) ( w \cS
-w)^{-1}, \label{Rew}
\end{eqnarray} 
where $u=\prod_{k=0}^{p} w_{k}$,
\end{Cor}
The Hamiltonian operators for equations (\ref{bogpro}) and (\ref{bogmod}) are given in
\cite{mr93c:58096}. They are
\begin{eqnarray*}
\cH_v&=&D_u^{-1} \cH {D_u^\dagger}^{-1}=v \left(
\left(\frac{\cS^{p+1}-1}{\cS^p-1}\right)\left( \frac{1-\cS^{-1}}{1-\cS^{-p}}\right)-
\left(\frac{\cS-1}{\cS^p-1}\right)\left( \frac{1-\cS^{-(p+1)}}{1-\cS^{-p}}\right)\right) v\\ 
&=&v   (\cS-1)\cS^{-1}  (\cS^{p+1}-1)(\cS^p-1)^{-1}  v
\end{eqnarray*}
and 
\begin{eqnarray*}
\cH_w&=&D_u^{-1} \cH
{D_u^\dagger}^{-1}=w\left(\frac{1-\cS^{-1}}{1-\cS^{-(p+1)})}-
\frac{\cS-1}{\cS^{p+1}-1)}\right) w \\ 
&=&w   (\cS-1) (\cS^p-1) (\cS^{p+1}-1)^{-1}  w,
\end{eqnarray*}
respectively. 

The equations (\ref{bogpro}) and (\ref{bogmod}) are also bi-Hamiltonian systems. They can be written as
\begin{eqnarray*}
v_t=\cH_v \delta_v  \prod_{k=0}^{p-1} v_{k}=\Re_v \cH_v \delta_v \frac{p \ln v }{p+1}
\end{eqnarray*}
and
\begin{eqnarray*}
w_t=\cH_w \delta_w  \prod_{k=0}^{p} w_{k}=\Re_w \cH_w \delta_w \ln  w
\end{eqnarray*}
respectively.

Here we make two remarks on the recursion operator $\Re$ in Theorem \ref{th1}:
\begin{enumerate}
\item Every pseudo-difference operator can be uniquely represented by its formal Laurent series \cite{mwx1}.
The highest order of the Laurent formal series of recursion operator $\Re$ in Theorem \ref{th1}
is $u \cS^p$ for all $p\in\bbbn$. Thus the fractional power of this pseudo-difference operator does 
not exist with coefficients in $u$ and its shifts, that is, we can not formally find an operator $Q$ such that $Q^p=\Re$.
\item  From the difference analog of the Adler Theorem \cite{mwx1, mr80i:58026}, it follows that
the residue and the logarithmic residue $\Re^l$, $l\geq 0$ are canonical conserved densities of the 
Narita-Itoh-Bogoyavlensky lattice.
\end{enumerate}

\section{Locality of symmetries}\label{sec4}
Recursion operators often have nonlocal terms. One important question
is whether the operator is guaranteed to generate local symmetries
starting from a proper seed. 
Sufficient conditions for weakly nonlocal \cite{MaN01} Nijenhuis
differential operators are formulated in \cite{mr1974732, serg5},
which are also valid for  weakly nonlocal Nijenhuis
difference operators \cite{mwx2}. This result is generalised to Nijenhuis
operators, which are the product of weakly nonlocal Hamiltonian 
and symplectic operators \cite{wang09}. 
However, the recursion operator given in Theorem \ref{th1} are not
weakly nonlocal. In this section, we are going to directly prove the
locality of symmetries by induction. 

To do so, we first introduce a family of homogeneous difference polynomials of
degree $l$ with respect to  the dependent  variable $u$ and its shifts
\begin{eqnarray}\label{sfunc}
\cP^{(l,k)}=\sum_{0\leq \lambda_{l-1}\leq \cdots \leq \lambda_0\leq k}
\left(\prod_{j=0}^{l-1} u_{\lambda_j+j p} \right),
\end{eqnarray}
where $k\geq 0, l\geq 1$ and $p\geq 1$ are all integers. In particular, for
any $p\in\bbbn$ we have
\begin{eqnarray}\label{sfuncp}
\cP^{(1,k)}= \sum_{j=0}^k u_j \quad \mbox{and} \quad \cP^{(l,0)}=u u_p u_{2 p}
\cdots u_{(l-1)p} \ .
\end{eqnarray}
Equation (\ref{bogadd}) can be written in terms of these polynomials as follows:
\begin{eqnarray}\label{bogfun}
u_t=u (\cS-\cS^{-p})\cP^{(1,p-1)} .
\end{eqnarray}

This family of polynomials was first defined in \cite{svin09}, where the author
amazingly gave the explicit expressions for the  hierarchy of symmetries of equation
(\ref{bogadd}) in terms of it. We are going to show that the recursion operator
(\ref{Re}) generates the same symmetries as in this paper modulo a sign. 

It is easy to see that the polynomials (\ref{sfunc}) possess the following
properties \cite{svin09}:
\begin{eqnarray}
&&\cP^{(l,k)}-\cP^{(l,k-1)}=u_k \cS^p (\cP^{(l-1,k)});\label{prop1}\\
&&\cP^{(l,k)}-\cS(\cP^{(l,k-1)})=u_{(l-1)p} \cP^{(l-1,k)}.\label{prop2}
\end{eqnarray}
These immediately lead to 
\begin{eqnarray}
&&(\cS-1) \cP^{(l,k)}=u_{k+1} \cS^{p+1} (\cP^{(l-1,k)})
-u_{(l-1)p} \cP^{(l-1,k)}.\label{prop3}
\end{eqnarray}
We now prove another important property.
\begin{Pro}\label{pro}
For all $l, p\in \bbbn$, we have
\begin{eqnarray}\label{imp}
(\cS^{p-i} u-u \cS^{-i})\cS^{-lp+i} \cP^{(l,(l+1)p-i)}= (\cS^{p-i}
u-u\cS^{-(i+1)} ) \cS^{-lp+i+1} \cP^{(l,(l+1)p-i-1)}, \ \ 0\leq i \leq p .
\end{eqnarray}
\end{Pro}
\noindent
{\bf Proof}. Let us compute the difference between the left-hand side and the
right-hand side of the identity (\ref{imp}) using the properties
(\ref{prop1}) and (\ref{prop2}):
\begin{eqnarray*} 
&&\quad (\cS^{p-i} u-u \cS^{-i})\cS^{-lp+i} \cP^{(l,(l+1)p-i)}- (\cS^{p-i}
u-u\cS^{-(i+1)} ) \cS^{-lp+i+1} \cP^{(l,(l+1)p-i-1)}\\
&&=u_{p-i} \cS^{-lp +p} \left( \cP^{(l,(l+1)p-i)} -\cS
\cP^{(l,(l+1)p-i-1)} \right)-u \cS^{-lp} \left(  \cP^{(l,(l+1)p-i)}- 
\cP^{(l,(l+1)p-i-1)}\right)\\
&&=u_{p-i} \cS^{-lp +p} \left( u_{(l-1)p} \cP^{(l-1,(l+1)p-i)} \right) -u
\cS^{-lp} \left( u_{(l+1)p-i}\cS^p \cP^{(l-1,(l+1)p-i)}\right)\\
&&=u_{p-i} u \cS^{-lp +p} \cP^{(l-1,(l+1)p-i)} -u u_{p-i}
\cS^{-lp+p}  \cP^{(l-1,(l+1)p-i)}=0 .
\end{eqnarray*}
We proved the statement.
\hfill $\blacksquare$

Notice that we can rewrite the recursion operator (\ref{Re}) in the form
 \begin{eqnarray}
&&\Re=u (\cS -\cS^{-p}) (\cS -1)^{-1} (\cS^{p} u-u\cS^{-1} ) \cdot
\prod_{i=1}^{\rightarrow{p-1}} 
(\cS^{p-i} u-u \cS^{-i})^{-1}(\cS^{p-i} u-u\cS^{-(i+1)} ) \nonumber\\
&&\quad \cdot ( u-u \cS^{-p})^{-1} \ . \label{rere}
\end{eqnarray}
Using it, we are able to prove the following result:
\begin{The} For Narita-Itoh-Bogoyavlensky lattice (\ref{bogadd}), starting
from the equation itself, its symmetries $Q^l=\Re^l(u_t)$ generated
by recursion operator (\ref{Re}) are local and
\begin{eqnarray}\label{syms}
Q^l=\Re^l(u_t)=u (1-\cS^{-(p+1)}) \cS^{1-lp} \cP^{(l+1,(l+1)p-1)}\quad \mbox{for
all} \quad   0\leq l\in\bbbz .
\end{eqnarray}
\end{The}
\noindent
{\bf Proof}. First the statement is clearly true for $l=0 $ from (\ref{bogfun}).
Assume the statement is true for $l-1\geq 0$. Let us compute the next symmetry
$Q^{l}$.
Taking $i=p$  in (\ref{imp}) and using the induction assumption, we have
\begin{eqnarray*}
( u-u \cS^{-p})\cS^{-lp+p} \cP^{(l,lp)}= (u-u\cS^{-(p+1)} ) \cS^{-lp+p+1}
\cP^{(l,lp-1)}=Q^{l-1}.
\end{eqnarray*}
Hence
\begin{eqnarray*}
&&Q^l=\Re (Q^{l-1})=u (\cS -\cS^{-p}) (\cS -1)^{-1} (\cS^{p} u-u\cS^{-1} )
\\
&&\quad \cdot\prod_{i=1}^{\rightarrow{p-1}} 
(\cS^{p-i} u-u \cS^{-i})^{-1}(\cS^{p-i} u-u\cS^{-(i+1)} ) \cdot \cS^{-lp+p}
\cP^{(l,lp)} ,
\end{eqnarray*}
where we used formula (\ref{rere}).
We now recursively apply formula (\ref{imp}) for $i$ from $p-1$ to $1$ and
obtain 
\begin{eqnarray*}
&&\quad Q^l=\Re (Q^{l-1})=u (\cS -\cS^{-p}) (\cS -1)^{-1} (\cS^{p} u-u\cS^{-1} )
\cS^{-lp+1} \cP^{(l,lp+p-1)} \\
&&=u (\cS -\cS^{-p}) (\cS -1)^{-1} \left(u_p \cS^{-lp+p+1}
\cP^{(l,lp+p-1)} -u \cS^{-lp} \cP^{(l,lp+p-1)} \right) \\
&&= u (1-\cS^{-(p+1)}) \cS^{1-lp} (\cS-1)^{-1}\left( u_{(l+1)p} \cS^{p+1} 
\cP^{(l,(l+1)p-1)} -u_{lp}  \cP^{(l,(l+1)p-1)} \right)\\
&&= u (1-\cS^{-(p+1)}) \cS^{1-lp} \cP^{(l+1,(l+1)p-1)} .
\end{eqnarray*}
Here we used formula (\ref{prop3}) for $l$ being $l+1$ and $k$ being
$(l+1)p-1$, that is,
\begin{eqnarray*}
&&(\cS-1) \cP^{(l+1,(l+1)p-1)}=u_{(l+1)p} \cS^{p+1} (\cP^{(l,(l+1)p-1)})
-u_{lp} \cP^{(l,(l+1)p-1)}.
\end{eqnarray*}
We completed the induction proof of the statement.
\hfill $\blacksquare$
\section{Discussion and further work}
Recently, a family of integrable lattice hierarchies associated with fractional
Lax operators was introduce by Adler and Postnikov \cite{adler1, adler2}. One simple
example is
\begin{eqnarray}\label{adler}
u_t=u^2 (u_p \cdots u_1-u_{-1}\cdots u_{-p})-u (u_{p-1}\cdots u_1-u_{-1}\cdots
u_{1-p})), \quad 2\leq p\in\bbbn ,
\end{eqnarray}
which is an integrable discretisation for the Sawada-Kotera
equation.  Notice that equation (\ref{adler}) is a combination of equations
(\ref{bogmod}) and (\ref{bogpro}) with different $p$. It can be considered as inhomogeneous 
generalisation of the Bogoyavlensky type lattices. To study its algebraic and geometric structures, 
we need first know the answers to  the Bogoyavlensky type lattices. This is the main motivation of
this paper.

As pointed by the authors of \cite{adler2}, an important open question of
this new family of integrable equations is to construct their Hamiltonian
structures. The present paper does not offer an immediate solution and further research is required.

\section*{Acknowledgement}
The author would like to thank A.V. Mikhailov and J.A. Sanders for useful discussions,
and gratefully acknowledges financial support through EPSRC grant EP/I038659/1.

\end{document}